# Out of the Loop: Problems in the development of next generation community networks

Murali Venkatesh

Ph.D., Associate Professor & Director
School of Information Studies
Community & Information Technology Institute (CITI)
4-279 Center for Science & Technology
Syracuse University
Syracuse, New York 13244-4100
(315) 443-4477 mvenkate@syr,edu

Donghee Shin

Doctoral Student
School of Information Studies
4-227 Center for Science & Technology
Syracuse University
Syracuse, New York 13244-4100
(315) 443-5618 dhsin@syr.edu

**Abstract.** Drawing on an ongoing longitudinal research study, we discuss problems in the development of five next generation community networking projects in central New York. The projects were funded under a state program to diffuse broadband technologies in economically depressed areas of the state. The networks are technologically complex and entail high costs for subscribers. The political economy of the development process has biased the subscriber base toward the resource rich and away from the resource poor, and toward tried-and-tested uses like Internet and intra-organizational connectivity and away from community-oriented uses. These trends raise troubling questions about network ontology and function, and about the relation between the network and its physical host community. The need for appropriate social policy, and new planning practices, is argued to effect desired change.

1. **Introduction**

Telecommunications design and deployment decisions are powerfully shaped by market forces. A political economic view holds that the forces "shaping the application and development of telecommunications are the political, economic, social and cultural dynamics of capitalism itself. Above all, especially in the neo-Marxist accounts, the development and application of telematics are seen to be driven by the imperative of maintaining capital accumulation of firms…" (Graham & Marvin, 1996, p. 95). A key premise of this view is that technology is not neutral. "Telecommunications are not neutral technologies. They are not equally amenable to all users…an inherent bias is already "locked in" to them through the network design process" (Gillespie, 1991, p. 225). This inherent
development of solutions in the public interest.

In this paper, we trace the development of five advanced technology community networks in central New York; we refer to these as next generation community networks. These networks are funded under a New York State program dedicated to diffusing advanced technology in economically depressed neighborhoods. Only public, non-profit and small business entities are eligible to subscribe to these networks; individuals and households, and state government organizations, are not eligible. Their technological complexity and subscription costs, both of which are significantly higher relative to dial-up community networks such as Freenets (see Serra, 2000, who views Freenets as examples of first generation community networks), have helped bias subscription toward resource rich and away from resource poor organizations. They threaten to polarize and exclude and, at this point in their development, favor the development of tried and tested uses of the network at the expense of innovative and locally responsive uses. Given their resource demands, the political economic view of telecommunications development applies even more urgently in the case of next generation community networks.

Their developmental trajectory raises basic questions for civil society. Are community network subscribers merely consumers or are they also citizens, with civic obligations to complement their rights as institutional members of the host community? As an instrument of civil society in the host community, such a network should mediate between a community's institutions and its residents. The evolution of the network, ideally, would be in response to local needs and values, shaped by social policy choices identified participatively in the locality. Although political economic forces have been prominent thus far in the developmental effort, effective social action to promote civil society ideals and resist the influence of commodifying forces will be critical in shaping the networks' evolution. We identify focal areas for social policy development and a new planning praxis in advanced telecommunications in the public interest.

**2. Background**

In 1995, as part of a settlement of a regulatory case before the NY State Public Service Commission, the incumbent local exchange carrier in the state (hereafter the carrier) committed $50 million to develop and deploy broadband telecommunications services in economically depressed communities across the state. The program[1] that came into being as a consequence established a competitive request for proposals process[2] to solicit proposals from eligible consortia of non-profit and public sector organizations -- city and county government agencies, K-12 schools and higher educational institutions, healthcare and human service agencies -- and small businesses. Organizations had to be located in or provide services to approved low-income zip code areas to be eligible for subsidies. State government agencies could not participate in the program. Two rounds of grants were awarded before the program concluded in 2000. In all, twenty-two projects were funded -- 14 urban/suburban, six rural, with two qualifying as urban/suburban/rural combination projects (Evaluation Report, 2001).

Subscribers would pay reduced monthly service charges and were eligible for some financial support toward computer and networking equipment at the user premise (customer premise equipment or CPE). Eighty per cent of every grant went back to the carrier for infrastructure and services development and deployment. The remainder could be applied toward CPE or training at the subscriber site. Grant funds could not be used to hire technical consultants or administrative staff, or to support applications development.

The goal of the program was to bring "advanced telecommunications services to economically disadvantaged areas of New York State that would not be available in the near future on account of limitations in the advanced telecommunications infrastructure and related equipment marketplace" (Evaluation Report, 2001, citing program documents). In other words, these areas would not have access to advanced services if it were left to market forces. Less formally but no less important as a goal, the program sought to encourage public sector, non-profits and small business entities to come together to solve problems local to the host community. Program selection committee members pointed to the strengthening of existing local relationships and forging of new ones in grantee communities as an important program goal. This idea of diverse institutions and interests coming together voluntarily for the public good was a recurrent theme in their conceptualization of a community. They viewed a program-funded network as an instrument to further cement social and institutional relationships, promote convergence and reciprocity, augment the community's ability cooperatively to address problems and opportunities, and to serve as a bridge across the Digital Divide. "Community networking", forging "community coalitions and partnerships", "finding common ground": these were some of the terms used by respondents to characterize "community". The network would amplify such sentiments: "Recall that, to be eligible, organizations had to connect to a common (technology) backbone: this reinforced the notion of community and information sharing". Cross-sectoral connectivity (i.e. where a K-12 school connected not just to other schools but, say, to the public safety building) was a "preferential factor" used to score proposals for awards under the program. "Applications (network uses) should cover a cross-section of the community" was offered as a condition for success. Program selection committee respondents' construal community", "community networking" and "community network" was "aspirational" rather than empirically grounded (see Kling & Courtright, forthcoming), and envisioned a desired end-state.

---

[1] The program required that approximately 80% of funds be allocated to urban-suburban regions of the state and the remaining 20% to rural areas. Additionally, the plan specified that approximately 80% of the funds were to be disseminated for network infrastructure 20% for customer premises equipment and associated training".

[2] The committee established to evaluate proposals and make comprised representatives from the carrier, the state Consumer Protection Board, the Public Utilities Law Project, the NY State Senate and Assembly, the NAACP, the NY State Board of Regents, the Empire State Development Corp., the NY State Departments of Health and State, and the Office of the Advocate for Persons with Disabilities).

Explaining the decision to use a competitive request for proposals process, a selection committee member said: "We were not interested in putting technology in place. (Proposers) had to make a strong case that the investment would make a difference to the community in economic and social terms". To be eligible for program subsidies, subscribers were required to connect to a common backbone. This way, subscribers could connect to the world at large through the Internet but would also be connected to other subscribers locally, in the host community. In the latter sense, a program-funded network was intended to serve as an inter-organizational infrastructure. To qualify for an award, proposals had to show sustainable, broad-based support for the project within the local community and lasting benefit from it, a well thought out, robust work plan, and sensitivity to the needs of citizens with disabilities.

**2. Background to research**

Our ongoing research program examines the process through which advanced community networks develop and evolve. As such, it is longitudinal in nature. We include planning, design and implementation under the term *development*. Our research is guided by two broad questions: *How do advanced community networks develop and evolve? What are the dynamics behind and consequences of technology choices?* We refer to these advanced technology networks as next generation community networks (NGCNs), and define NGCNs below.

In this paper, we trace the development of five next generation community networks in six communities in central New York (Table 1). These projects – one urban/suburban and four rural – were awarded over $11 million in funding in the program's second round.

| Project | Program Category | Grant Amount |
|---|---|---|
| Central New York State (one project, two communities) | Urban/Suburban | $3.7m (revised) |
| Western New York State | Rural | $2.9m |
| Southern New York State | Rural | $1.5m |
| Northern New York State | Rural | $1.5m |
| Eastern New York State | Rural | $1.5m |
| Number of projects | 5 | |
| Number of communities | 6 | |
| Total Program funding for five projects | | $11.1m |

**Table 1 Communities and Projects Surveyed in Research**

The program identified certain zip codes in the carrier's service area as economically depressed. In these zip codes, the median household income was below 75% of the statewide median, as determined by program authorities from 1990 census figures. Additionally, in several of these areas, the percentage of households without telephone service was determined to be at least 50% above the statewide average.

We refer the networks being developed in these communities as community networks in the empirically grounded sense that their physical context is defined by economically depressed zip code areas in the host communities. We tracked the work of volunteer planners in these communities for this research; additionally, we also gathered data from carrier design staff and program selection committee members. We have tracked these communities since 06/1996, starting with NGCN planning process in our community. Data on NGCN design (technical specification) was collected in two phases (06/1998 to 05/2000). In Phase 1, we interviewed planners from the six communities to document their view of the design process. In Phase 2, which concluded in May 2000, data was collected from program selection committee members and carrier staff from engineering, marketing and sales functions.

The groups we call "planners" typically were white collar residents of the host communities. Some had knowledge of technology; most did not. None had a background or experience in urban planning. Most represented public institutions or non-profits in the community; others were private citizens with an interest in the project. The groups formed voluntarily and were self-selected, with the initial impetus to organize coming from the carrier, and came into being in response to the funding opportunity. By and large, the groups shared the aspirations of the program selection committee for the network. These groups are referred to as the "planners" in this paper, as they were responsible for securing the grant and represented the community's interest in developing the network.

The first author recently completed his biannual survey/interview of project principals at all five project sites for an update on heir development and governance. This data is referenced in this analysis as well.

**3. Advanced Technology Community Network, or NGCN**

As a broadband network infrastructure, the advanced technology network described here is significantly more complex compared to dial-up community networks (powered by the telephone modem) of the first generation (see Serra, 2000). We refer to such a network as a next generation community network – an NGCN. Our characterization is limited to the projects described here and should not be taken to represent advanced community networks in general.

Tariffed telecommunications services (i.e. widely available services governed by rates and rules subject to review by the Public Service Commission) were ineligible for subsidies in the program's second round. The carrier would offer non-tariffed broadband services as a limited service offering (LSO) to subscribers. The fact that the services were not tariffed rendered them discontinuous with existing infrastructures and contracting arrangements for subscribers; carrier design staff were also impeded to an extent by the relative novelty of these services.

The community networks examined share the following technical and related features, which constitute our definition of an NGCN. In the remainder of the paper, these networks are referred to as NGCNs.

- *High technological complexity*: The NGCN is a technologically complex, broadband, open, multi-service and multi-layered community intranet with high-speed Internet access. It can support data transfer at high to very high rates (1.5 Mbps to 1 Gbps range). Gigabit Ethernet and Asynchronous Transfer Mode (ATM) cell relay service are the services available to subscribers. As a broadband environment, the NGCN can support delivery of high-touch services (high-bandwidth services with rich media content, like video-conferencing) while also supporting conventional uses (information exchange and email). The NGCN is not merely a logically carved-out and locally focused segment of the Internet (see Serra, 2000) but inverts the paradigm: it is a hardwired, broadband, intranet (access restricted to eligible organizations) that also connects subscribers to the Internet.

- *Costly and complex applications development*: Developing applications for the NGCN is resource-intensive and technically complex. For example, the community library system would like to digitize its holdings and place them online through a portal accessible over the NGCN and the Internet. The design and development of such a multi-media application with adequate security and service quality assurance calls for complex skill sets and significant resources.

- *Complex services contracting*: Contracting for NGCN services can be complex. Figure 1 analyzes the services needs and contracting challenges from the subscriber's viewpoint. As their needs evolve, subscribers will be faced with in-source vs. buy decisions on a broad range of services across the architecture, potentially from multiple service providers in a competitive marketplace. While this will benefit subscribers, it will also complicate contracting.

- *High subscription costs*: Subsidized subscription charges range from approximately $300 a month for Gigabit Ethernet to 15 times that for high-end ATM service. Subscribers pay extra for Internet access, and an additional charge for network management and related services. Additional resources would be required if software applications had to be developed or acquired. In 1998, relatively cheaper DSL was eligible for program subsidies. Many non-profits were interested in DSL for its affordability. However, it subsequently became a tariffed service and became ineligible. The technology committee suggested variations on the tariffed DSL service that might be eligible for subsidies, but the carrier was not open to exploring these options.

**Application Layer Services:** Network management**,** application troubleshooting (help-desk functions)

**Application Implementation Services:** Developing and implementing World Wide Web-enabled applications, security and privacy, integration of data conferencing with World Wide Web applications and videoconferencing, MPEG/JPEG conversion

**Value-added Internet (IP ) Services**: Multicasting, video bridging, switching and gateway services, Internet (IP) telephony

**Basic Internet (IP) Services**: Management of standard IP services up to and within participant site

**Layer 2 Services:** Gigabit Ethernet, Cell Relay/ATM, Frame-to-Cell Conversion, LAN Emulation, integration with existing LANs, (re)designing LANs for multi-media applications sharing

**Physical Layer Services:** DSn over copper and fiber, SONET.

**Figure 1**: Services Reference Architecture (SRA), showing a sampling of technical support services needs in the NGCN
*Copyright Community & Information Technology Institute (CITI), 1999*

### 4. Findings: Political economy of NGCN Development

Recall that the NGCNs were designed to serve as an interorganizational system infrastructure. It is not surprising then that their development has been powerfully influenced by the economic interests of large public institutions. Proponents of the system resource view argue that access to resources and power are basic goals in organizations and in inter-organizational relations (Benson, 1975). "Political economy" links a polity – the political environment -- with an economy (Wamsley & Zald, 1973). The polity refers to power systems and structures internal and external to the firm and the ends or values they are used to achieve. A firm's economy, similarly, has internal and external connotations. Intern
to the resources a firm commands and how they are allocated; externally, it refers to the framework – the rules, mechanisms – that govern its relations with the larger economic environment. The polity and economy are interrelated: internally, power systems affect resource allocation and division of labor, and access to resources provides power. Externally, firms that can command resources are seen as powerful by other entities in the environment.

Currently, there are very few small non-profit or small business entities that are NGCN subscribers in the communities surveyed; large public institutions dominate the subscription lists. For example, the Syracuse NGCN currently has 12 subscribers, with a total of 204 endpoints  (connections) among them. The list includes city and county government agencies, K-12 public schools, and large non-profits like Syracuse University and a major hospital. Small non-profits in the social sector (independent of government) and small businesses are notably absent from the list. These entities often provide key services in low-income neighborhoods. Eight such entities, as part of a separate coalition outside the program, have negotiated subsidized digital subscriber line (DSL) service with a competitive local carrier for Internet access and internal connectivity. The coalition is linked to the program-funded network.

A recent report by an independent consultant, which reviewed all 22 projects funded under the program, concluded: "those institutions already involved in technology and advanced technology such as BOCES (a county-level technology resource organization for K-12 schools), community colleges and hospitals were predisposed to or ready to take full advantage of the program" (Evaluation Report, 2001, p. 29). Generally speaking, public institutions like these have the resources – money, people, technology -- to support use of advanced technology and are significantly better off in this respect relative to the average non-profit and small business. They also have the power – the clout -- in communal deliberations to sway decisions. The NGCN development process, consequently, has been shaped by their economic interests. This is not to say that they exercised power consciously to manipulate project goals. They were seen by planners as vital "to get the network up and running" by being early adopters, and their influence stemmed from their putative role as guarantors of the project going forward. As early adopters, their interests were considered crucial to the community having an NGCN at all.

At one point during the design, there was an implied threat from program authorities to revoke the grant if not enough institutions signed on as subscribers. As an expedient response, a Mall model of diffusion was adopted by planners. The argument was as follows: get anchor tenants (early adopters) to sign on and then worry about recruiting others. The Mall model was useful in starting up the network. But in the absence of a covenant requiring subscribers to assist the resource-poor, the model likely will have limited impact as a broad diffusion mechanism. Market forces will continue to apply to narrow the subscriber base to those that can afford to pay. The more the use of a network "depends on social organization and mobilization of significant resources, the more it will tend to be controlled by those who are already organized and well-off" (Calhoun, 1998, p. 383). This assertion applies with even greater urgency to NGCNs because of their resource demands on subscribers.

The Mall model was also attractive to the carrier. The carrier operated in a highly competitive market, with competing local exchange carriers vying for the same large accounts. Locking in large accounts early was an outcome attractive to the carrier.

High subscription costs and the NGCNs' technological complexity raise troubling questions about access equity and distributional justice. Continuance of the program subsidies to address the needs of community organizations was proposed recently before the Public Service Commission:

> "Because of the funding constraints which they experience…community service organizations in low income areas are often unable to incorporate…advanced technologies into their operations…By focusing the available rate reductions on this sector of service providers in low income communities …(the continuance of the program) will help bring these customers more directly into the digital economy and…bridge the digital divide" (Public Utility Law Project and NY State Community Action Association, Comments filed with the PSC, 2000, pp. 3-4).

The document continues:

> "The non-profit organizations eligible for the reduced rates would include community action agencies, day care centers, Head Start providers, community centers, community health care clinics, legal aid offices, and other service providers located within low income communities. By reducing broadband rates for these organizations, real assistance is provided to the organizations which create the social infrastructure on which low income communities depend and, therefore, to the low income families and households which make up that community" (p. 3).

Importantly, besides rate reductions (subsidized network access charges), the document seeks assistance to cover technical support and equipment purchase at the customer premise.

The high subscription costs have favored tried-and-tested uses of the NGCN. Subscribers have replaced existing service contracts for Internet access and internal connectivity with the subsidized services, saving significant amounts of money. However, such uses balkanize the NGCN. Subscribers are connected to the world and to their branch offices, not necessarily to one another or to community residents. Internet access and internal connectivity alone do not call for a community network. Applications that deliver information or services to community residents are more fitting uses, but these can be costly and complex to develop. Most of the 22 projects examined, including all five projects surveyed here, cited resource constraints as the most serious challenge to ongoing development of the NGCN (Evaluation Report, 2001). Community-oriented uses of the NGCN, of whatever kind, may never materialize due to resource constraints.

Recall that the program required that the NGCN be based in and advance the interests of the poorer sections in the host community, and planners' aspirations for it were consistent with this focus on locality. Descriptive, non-evaluative definitions of the term "community" start with the notion of locality: a community, in its basic sense, is tied to a particular physical locale (see Khatchadourian, 1999). Accordingly, it has been argued that a community network should be "as grounded as possible by primarily gearing toward local users" (Aurigi & Graham, 2000, p. 498). What, then, is the relationship between the NGCNs and their physical host communities? Our research shows that this relationship is tenuous at best.

Take, for example, the NGCN in Syracuse. The NGCN, it was hoped, would foster the economic and social revitalization of Syracuse. But it was not seen by the urban planning establishment as part of the city's present social reality or its future. City and county planners played only a small, non-substantive role in the development process: they generated GIS maps on eligible zip codes at the project's outset. They viewed the project as "too esoteric" and as "not impacting the grassroots", alluding to the NGCN's technological complexity and high subscription costs. Elected officers and the mass media played no part in the project (one elected official, the county sheriff, participated briefly early on in the project; see Castells, 1996, and Mino, 2000, on need for strong political and media participation in such community initiatives). This meant that the NGCN was not part of any transforming, inclusive, strategic vision for the city, nor was it viewed in relation to tactical urban planning initiatives or goals. In other words, the NGCN was not part of anything larger than itself. A master plan for the city is currently in preparation by the Mayor's office and is expected to reference the NGCN. However, the civic vision was unavailable to inform, or be informed by, the NGCN in its crucial formative stages, when questions of equity and network ontology and function could have been debated most profitably (Aurigi, 2000, notes a similar disconnect between urban planners and "digital city" projects in Europe). Consider the following:

➢ The over-65 population in the four county region, including Syracuse, rose by over 4% since 1990 (Syracuse Post Standard, June 15, 2001). This increase is stretching the capacity of nursing homes and retirement communities, and the situation is expected to get worse. Lack of transportation options is also an issue, particularly in suburban and rural areas, for the growing number of elderly in the region.

➢ Syracuse City's racial makeup grew more varied between 1990 and 2000. The black, Asian and Hispanic populations grew, and that of whites declined. Non-Hispanic whites make up 62% of the city's population, but constitute only 42% of the under 17 population. By 2010, the white population is predicted to be under 50% in the city (Syracuse Post Standard, March 26, 2001).

There are no service applications, and no proposals for such applications, targeting the needs of the elderly or the city's many racial and ethnic groups. Currently the poor, whose numbers and spatial concentrations provided the physical and moral context to the project, are not covered either. Diverse populations and minority groups in the K-12 schools have faster access to the Internet via the NGCN, but it is unclear how adult populations stand to benefit from such uses. There are no proposals to use the NGCN to combat urgent social problems such as adolescent pregnancy and infant mortality (in the late 90s, the county's infant mortality rate was the nation's highest). Socially responsible and community-oriented applications may yet develop as the NGCN evolves, but will be dependent on the vagaries of external funding. Meanwhile, the community at large may see little benefit from the grant.

This raises a fundamental question: Do subscribers have obligations to the physical host community? Should subscribers be viewed as (institutional) citizens or merely as consumers? The right of an eligible agency to program subsidies was recognized among planners and community institutions and came up in discussions of NGCN development policy. The corresponding notion of subscriber obligations to the host community came up only once. A city agency representative argued: "Since large players are getting such a good deal (from subsidized services), couldn't we require them to provide resources and services to smaller agencies? Large players have to see themselves as resource providers. This is part of their responsibility". Arguably, resource-rich subscribers could have helped, or could have been required to help, to promote the project's social goals, thus reducing reliance on external funding for the development of the NGCN. In the dialectical sense, requiring obligations of subscribers to meet the needs of the host community is a change mechanism that might have worked in this context. But it was never pursued.

Citizenship in a polity confers rights and requires obligations of rights holders. It is central to the ideal of civil society – which is founded on reciprocity and convergence of interests and values in the community. Aspirations consistent with those of civil society were an integral part of the morphology of the NGCNs, as both planners and selection committee members agreed. Mandelbaum (2000) notes: "the beginning of the search for a craft of community design is the notion of membership. Members are bound to one another by a web of rights and obligations" (p. 10). To the extent that it seeks to further the community it is a part of, an NGCN and its subscribers cannot escape the reciprocity of obligations (if only in the contractual sense), especially when the network is funded with rate payers' monies (public funds). Surprisingly, by-laws for NGCN governance in the projects surveyed make no mention of subscribers' obligations, of any kind, to any quarter. Subscribers were expected/encouraged, but not required, to "support projects that will benefit the underserved in the community", according to the project leadership responding to the first author's recent biannual survey. In the follow-up interview, respondents talked about their preoccupation with "getting the network off the ground". One of them implied that subscriber obligations would follow:

> "I hadn't thought of that (subscriber obligations). Maybe we could have language to that effect in the by-laws or in the governance somewhere. Currently, each of the large agencies has its own contract with the carrier. But we will have to pull together the board of directors…to begin to think collectively as a community, and think of obligations".

At present, subscribers have no obligations to be citizens, merely to be consumers. This emphasis is consistent with the Mall model of development and was, according to this respondent, occasioned by its imperative: to "get the network off the ground". This could only be done by signing up early adopters. It is not clear how a generalized expectation or non-specific obligation outside the by-laws, as in "support projects that will benefit the underserved in the community", can be enforced.

In the realm of citizenship, as O'Neill (1998) has argued, "it is the discourse of obligations rather than that of rights that is the primary vocabulary of action-centered ethics. When we discuss obligations, our direct concern is with what should be done" (p. 105). Increasingly, the focus is on our obligation as citizens to meet the needs of others in the community (see Brock, 1998). Specialized needs – such as healthcare for jail inmates or the elderly or sign language services for the hearing-impaired – constitute a large part of "our most urgent needs…And a large proportion of the cases in which people are not currently satisfying their own needs are cases in which the social structures in which people find themselves provide them with no feasible opportunity for doing so" (Baker & Jones, 1998, p.220). Community networks capable of delivering services can extend the reach of, and accessibility to, the social support infrastructure available in the community for need fulfillment; they can help adapt these social structures to the needs of residents with impairments and disabilities. As an extender of the support infrastructure, community networks can alter the social structure of choice and make it more convenient for the needy to avail of ameliorative services (see Baker & Jones, 1998). The NGCNs surveyed are fully capable of supporting delivery of both informational and high-touch services.

It is in the enhancement of the quality of day-to-day living, particularly for the less privileged, that NGCNs must find their true role and justification. Such a role is in fact emphasized in current visioning of electronic government or a teleserviced "e-polity" where a broad range of public services are delivered over telecommunications connections (see, for example, New York State Lt. Governor's Taskforce on Quality Communities, 2000). Informational services may be differentiated from high-touch services. The first delivers information to and may collect information from residents through electronic forms. For example, someone who has just moved from welfare to work should be able to apply for child-care vouchers, food stamps and other benefits online. The biggest advantage of online access to services is flexibility: users can go online at their convenience for asynchronous access to information and services. The ability to time-shift is critically important to individuals in low-wage jobs, which are far less flexible in allowing workers time off relative to white collar jobs (Wall Street Journal, August 27, 2001). As noted earlier, there are no plans at present to offer such services over the NGCNs surveyed.

The second enables live, synchronous, person-to-person interactive transactions, often video-based, such as telemedicine. Informational services can be (and indeed routinely are) delivered over dial-up technologies; one would not need broadband to deliver such services. High-touch services, on the other hand, are best delivered over broadband. For a person who urgently needs, say, a medical diagnosis, a network that delivers such a service is more valuable than one providing access to information alone (see Ishida, 2000). Considering that their high subscription costs has put the NGCN out of reach of community organizations and their clientele, online delivery of public services may be the only way (at least in the near term) to spread the benefits of expensive public investments in socially responsive ways. How the digital intersects with day-to-day life in the physical host community is a fundamental question for an NGCN. If used well, an NGCN can enrich and expand the interface between a community's residents and its social support infrastructures for need fulfillment and, more generally, for what Friedmann and Douglas (1998) call "human flourishing". But whether the digital intersects with life in the community at all will depend on whether subscribers are willing to be citizens first.

## 5. Conclusion

The foregoing analysis was not intended to assign blame to planners or the carrier for the way the NGCNs have turned out. The objective was to show the bias inherent in telecommunications (Graham & Marvin, 1996) and demonstrate how such a bias, which tends to favor the resource rich, may operate less equivocally in the case of advanced technologies. Furthermore, in this case, the bias was in play despite the community centered aspirations for the NGCN voiced by planners early on.

Telecommunications design and deployment decisions, in general, are powerfully shaped by market forces. Community networks have the opportunity to follow a different path, to envision new styles of community, a more equitable social order. A social activist view believes in the possibility of reconstruction and change in social systems. However, fundamental to this conception of social construction of technological outcomes is human agency. As Graham and Marvin (1996) have argued, a purely political economic analysis tends to view developmental processes in telecommunications as determined by "abstract and macro-level" power systems. We have tried to avoid this pitfall. While such forces are doubtless important, and have been key in the NGCN's early development, we believe that human agency can influence how it *evolves*. Combining the political economic and social activist views permits consideration of the developmental process at both societal and social interaction levels; such a combined view is emphatically contra determinism – of the technological or political economic variety. Apropos urban development in general, Imrie, Pinch and Boyle (1996) note the relevance of the political economic view and of social activism:

> "variants of political economy recognize …the need for taking a stand, for (re)engaging politically, for subjecting the (academic) analytical gaze to the materialities of cities…It is to be hoped that the (re)activation of political economy will become the core concern to those within urban studies" (p. 1260).

A social activist agenda would include a multi-level effort as well: at the macro level, there is the need for social policy fostering the delivery of teleservices over the NGCN as a way to reduce the effects of social exclusion; at the micro-level, the planner of advanced telecommunications solutions in the public interest has to be sensitive to the dynamics of development of the host community. In other words, the host community provides the necessary context within which the development of the NGCN should be located and analyzed.

Using the NGCN as a teleservice delivery mechanism can cut social isolation for certain populations. We noted above that the increase in the over-65 population in central New York has highlighted the lack of transportation options for the elderly in certain areas. Availability of transportation and convenient access to information and services are key elements of social inclusion in urban developmental analyses. Remote delivery of information and services using telecommunications to combat such problems makes a lot of sense, but is not common:

> "there are…major social, economic and political problems for which electronic solutions offer considerable potential, and which are as yet receiving less consideration. …electronic solutions can improve access to public services…the literature on contemporary urban environments has, for the most part, paid little attention to the development of such electronic solutions" (Turner, Holmes & Hodgson, 2000, p. 1724).

A community may be defined as a structure of social relations that ties its residents to one another and to its institutions. These social relations may serve a broad range of purposes, ranging from need fulfillment to protection from external or internal threats to fostering a sense of community. In terms of need fulfillment through access to public services, NGCNs can effectively disembed the social support infrastructure in a community from its spatial infrastructure (roads, transportation, parking), helping to develop a purely "relational" community connecting the needy with teleservice providers. To assign such a role to the NGCN is not to reduce it to a utilitarian function but to use it to actively cut social isolation and exclusion and make its benefits available to marginalized populations in meaningful ways. To use it thus is to use it as a vehicle to deliver stated social policy objectives, such as improved care of the elderly. However, as Nelson (1962) astutely noted, "there are orders of needs or values which vary with structure (a community's structure of relations). These orders may be called the hierarchy of values" (p.21). At present, the delivery of teleservices or the reduction of social isolation for such populations are not high on the list of priorities of the NGCNs surveyed. A deliberate aligning of the NGCN development effort with that of the host community, so notably absent in the communities surveyed, might have helped synchronize the two value hierarchies.

The use of telecommunications networks to deliver public services is relatively uncommon (see Turner, Holmes & Hodgson, 2000, on trends in the UK). A recent survey of networks in US communities suggests a similar trend (see Aurigi & Stephen, 2000). As we noted earlier, to use it as a mechanism to deliver urgently needed public services would be to spread the NGCN's benefits broadly in the host community. As has been argued (Turner, Holmes & Hodgson, 2000), we need a social technology policy to ensure that such uses are part of the thinking on NGCNs. The technological tools exist to deliver informational and high touch services remotely; we need to mobilize the will and develop the requisite policy instruments to effect the desired ends.

Social policy in this instance can serve to reorient the NGCN toward local needs and toward the NGCN's obligations to meeting these local needs. A focus on locality would dictate that each NGCN develops with reference to the needs and interests, and the hierarchy of values prominent in the community it is designed to serve. The determination of such needs, interests and values will ideally be the work of an empowered cadre of community residents. Lessons learned from the Acts of Enclosure in 18th and 19th century England are instructive here. These laws helped the powerful new (private) proprietors to bypass the local in favor of remote control and management. These laws, Smith (2001) explains

> "transferred open spaces into discrete areas bounded by fence or hedgerows, from places where social and economic practices were determined by tradition and local agreement to private domains for the privileged to own and exploit…the common lands were usually locally run with minimal outside interference, and for this reason they varied considerably, reflecting local environmental and sociopolitical histories" (p. 155).

Graham and Marvin (1996) argued that a "paradigm crisis" threatened urban studies and planning:

> "…many urban analysts and policy-makers still see cities through analytical lenses which…have less and less to do with the real dynamics of telecommunications-based urban development" (p. 48).

They call for education in and improved understanding of telecommunications in urban development. We see the need for the opposite to occur as well: community networking is imperilled if it is not informed by the dynamics of community development. During the planning process, planners had assumed that the city or community planning establishment would represent *community* needs. This turned out to be a mistake: the planning establishment did not play a major role in NGCN development, as noted earlier, which meant that the community's needs went under-represented. The new planner would be well enough informed about the *community's* needs to work with the planning establishment to represent them, or represent them herself, in the developmental process.

Community network development in the public interest cannot be removed from the concerns of the host community. In bridging this distance, community network planners can learn much from the evolution of urban planning to its contemporary emphasis on radical practice:

> "Radical practices emerge from experience with and a critique of existing unequal relations and distributions of power, opportunity and resources. The goal of these practices is to work for structural transformation of these systemic inequalities and, in the process, to empower those who have been systematically disempowered (Sandercock, 1998, p. 176).

What then might be the role of the new, radicalized planner in community network evolution? We see three ways into the new praxis:

➢ Despite their sympathetic view of planners' aspirations, the carrier design staff were more attuned to a technical rational view of network development, which seeks to optimize outcomes on conventional design criteria such as network performance and efficiency. The new planner has to ensure that the question of ends is not lost sight of in the shuffle of purely technical rational preoccupations. Design choices can divide and balkanize and have the force of implicit policy (Guthrie & Dutton, 1992). The planner has to resist and reorient such trajectories. Technical criteria are critical in design, but so are the larger issues – the ontology, the ends – that design must serve. In her new role, the planner would use means and viewpoints considered vital in current urban planning practice – the ability to listen and question in order to learn (Sandercock, 1998), and a post-modern sensibility (Graham & Marvin, 1996), to keep the focus on the *community's* interests and needs.

➢ The new planner would combine technical and contextualized knowledge with an understanding of "radical" developmental methods to empower and invite participation from diverse publics. Given the high knowledge demands of NGCN development, the new planner has to fulfill the role of an expert advocate in the network infrastructure design process. Her role as expert advocate is indispensable, and unavoidable, given the technical nature of the process. However, in terms of ideas for network use, the community (or *communities*) should be in charge, with the planner as an ally. Radical methods like applications prototyping can be used to highlight urgent needs in the community and to influence the agenda for network uses around such urgent needs. Working with a mediating institution would be especially beneficial if prototyping is involved, on account of the resources that would be needed. There are other methods to build a constituency. For example, the new planner can publicize exemplars of innovative uses of technology for pro-social ends to develop grassroots momentum.

➢ Political economic forces cannot be wished away. Radical praxis has to acknowledge the pragmatics of NGCN development. The new planner would be an activist willing to take sides and work through politics and conflict to recenter marginalized interests. In this, she has to work to convert community-centered aspirations into explicit social policy instruments to effect desired change. As we showed, this conversion has not occurred with the NGCN. It is vitally important that subscribers be made aware of their obligations to the host community. This must occur in a timely fashion, in order to shape the evolution of the NGCN such that it intersects substantively and meaningfully with community life.

**References**

Aurigi, A. (2000). Digital City or Urban Simulator?, Digital Cities, Lecture Notes in Computer Science, 1765. Berlin: Springer-Verlag, pp. 33-44

Aurigi, A., and Graham, S. (2000). Cyberspace and the City, The Virtual City in Europe. Companion to The City. Oxford: Blackwell, pp. 489-502

Baker, J. and Jones, C. (1998). Responsibility for Needs. In G. Brock (Ed.), Necessary Goods: Our Responsibilities To Meet Others' Needs. Lanham, MD: Rowman & Littlefield Publishers, pp 219-232.

Brock, G. (1998). Introduction. In G. Brock (Ed.), Necessary Goods: Our Responsibilities to Meet Others' Needs. Lanham, MD: Rowman & Littlefield Publishers, pp. 1-18

Calhoun, C. (1998). Community without Propinquity Revisited: Communications Technology and the Transformation of the Urban Public Sphere. Sociological Inquiry, 68 (3), pp. 373-397.

Castells, M. (1996). The Information Age: Economy, Society and Culture. The Rise of Network Society, Vol. I. Oxford: Blackwell.

Castells, M. (1985). High Technology, Economic Restructuring, and the Urban-regional Process in the United States. In M. Castells (Ed.), High technology, space, and society. Beverly Hills, CA: Sage Publications, pp. 1-40

Castells, M. (1999). The Informational City is a Dual City: Can it be Reversed? In D.A. Schon, B. Sanyal & W. J. Mitchell (Eds.), High technology and low-income communities: Prospects for the positive use of advanced information technology. Cambridge, MA: The MIT Press, pp. 25-41.

Evaluation Report.(2000). New York State Advanced Telecommunications Project: Diffusion Fund Program. White Plains, NY: Magi Educational Services (mimeo).

Friedmann, J. and Douglass, M. (1998). Editors' introduction. In M. Douglass & J. Friedmann (Eds.), Cities for citizens: Planning and the rise of civil society in a global age. Chichester, UK: John Wiley, pp. 1-6.

Gillespie, A. (1991). Advanced Communications Networks, Territorial Integration, and Local Development, in R. Camagni (Ed.), Innovation Networks. London, UK: Bellhaven, pp. 214-229.

Graham, S. and Marvin, S. (1996). Telecommunications and The City: Electronic Spaces, Urban Places. London, UK: Routledge.

Guthrie, K. and Dutton, W. (1992). The Politics of Citizen Access Technology: The Development of Public Information Utilities In Four Cities. Policy Studies Journal. 20 (4), pp. 574-597

Imrie, R., Pinch, S., and Boyle, M. (1996). Identities, citizenship and power in the cities. Urban Studies, 33 (8), pp. 1255-1261.

Ishida, T. (2000). Understanding Digital Cities, Digital Cities, Lecture Notes in Computer Science, 1765. Berlin: Springer-Verlag, pp. 7-17

Khatchadourian, H. (1999). Community and Communitarianism. New York, NY: Peter Lang.

Kling, R. and Courtright, C. (forthcoming). Group Behavior and Learning in Electronic Forums: A Socio-Technical Approach. To appear in: S. Barab, R. Kling and J. Gray (Eds.), Building online communities in the service of learning. Cambridge University Press.

Mandelbaum, S.J. (2000). Open Moral Communities. Cambridge, MA: The MIT Press, (2000)

O'Neill, O. (1998). Rights, Obligations, and Needs. In G. Brock (Ed.), Necessary Goods: Our Responsibilities to Meet Others' Needs. Lanham, MD: Rowman & Littlefield Publishers, pp. 95-112

Public Utility Law Project and New York State Community Action Association, (July,2000). Comments filed with the Public Service Commission

Sandercock, L. (1998). The Death of Modernist Planning: Radical Praxis for a Postmodern Age. In M. Douglass & J. Friedmann (Eds.), Cities for Citizens: Planning and the Rise of Civil Society in a Global Age. Chichester, UK: John Wiley, pp. 163-184

Serra, A. (2000). Next Generation Community Networking: Futures for Digital Cities, Digital Cities, Lecture Notes in Computer Science, 1765. Berlin: Springer-Verlag, pp. 45-57

Smith, M. (2001). An ethics of place: Radical ecology, postmodernity, and social theory. Albany, NY: State University of New York Press.

Turner, J., Holmes, L., and Hodgson, F.C. (2000). Intelligent urban development: An introduction to a participatory approach. Urban Studies, 37 (10), pp. 1723-1734.